\documentclass[a4paper,fleqn,usenatbib]{mnras/mnras}
\pdfoutput=1
\usepackage{newtxtext,newtxmath}

\usepackage[T1]{fontenc}
\usepackage{ae,aecompl}

\usepackage{graphicx}	
\usepackage{amsmath}	
\usepackage{amssymb}	
\usepackage[utf8]{inputenc}
\usepackage[flushleft]{threeparttable} 
\usepackage{hyperref} 

\title[A lensed quasar discovered with PanSTARRS1 and Gaia]{The Discovery of a Five-Image Lensed Quasar at z = 3.34 using PanSTARRS1 and Gaia} 

\author[F. Ostrovski et al.]
{\parbox{\textwidth}
{Fernanda Ostrovski$^{1,2}$\thanks{E-mail: fernanda.ostrovski@ufrgs.br},
Cameron A. Lemon$^{2,3}$,
Matthew W. Auger$^{2}$,
Richard G. McMahon$^{2,3}$,
Christopher D. Fassnacht$^{4}$,
Geoff C.-F. Chen$^{4,5}$,
Andrew J. Connolly$^{6}$,
Sergey E. Koposov$^{7,2}$,
Estelle Pons$^{2,3}$,
Sophie L. Reed$^{2,3}$,
Cristian E. Rusu$^{4,8}$
}
\\~\\
{\small
$^{1}$Departamento de Astronomia, Instituto de F\'{i}sica da Universidade Federal do Rio Grande do Sul, 91501-970, Porto Alegre, Brazil}\\
$^{2}$Institute of Astronomy, University of Cambridge, Madingley Road, Cambridge CB3 0HA, UK\\
$^{3}$Kavli Institute for Cosmology, University of Cambridge, Madingley Road, Cambridge CB3 0HA, UK\\
$^{4}$Department of Physics, University of California Davis, 1 Shields Avenue, Davis, CA 95616, USA\\
$^{5}$Institute of Astronomy and Astrophysics, Academia Sinica, P.O. Box 23-141, Taipei 10617, Taiwan\\
$^{6}$Department of Astronomy, University of Washington, Seattle, WA 98195, USA\\
$^{7}$McWilliams Center for Cosmology, Department of Physics, Carnegie Mellon University, 5000 Forbes Avenue, Pittsburgh, PA 15213, USA\\
$^{8}$Subaru Telescope, National Astronomical Observatory of Japan, 650 N Aohoku Pl, Hilo, HI 96720\\
}

\date{Accepted XXX. Received YYY; in original form ZZZ}

\pubyear{2017}

\begin{document}
\label{firstpage}
\pagerange{\pageref{firstpage}--\pageref{lastpage}}
\maketitle

\begin{abstract}
We report the discovery, spectroscopic confirmation, and mass modelling of the gravitationally lensed quasar system PS J0630$-$1201. The lens was discovered by matching a photometric quasar catalogue compiled from Pan-STARRS1 and WISE photometry to the Gaia DR1 catalogue, exploiting the high spatial resolution of the latter (FWHM $\sim 0\farcs1$) to identify the three brightest component of the lensed quasar system. Follow-up spectroscopic observations with the WHT confirm the multiple objects are quasars at redshift $z_{q}=3.34$. Further follow-up with Keck AO high-resolution imaging reveals that the system is composed of two lensing galaxies and the quasar is lensed into a $\sim$2\farcs8 separation four-image cusp configuration with a fifth image clearly visible, and a 1\farcs0 arc due to the lensed quasar host galaxy. The system is well-modelled with two singular isothermal ellipsoids, reproducing the position of the fifth image. We discuss future prospects for measuring time delays between the images and constraining any offset between mass and light using the faintly detected Einstein arcs associated with the quasar host galaxy.
\end{abstract}




\section{Introduction}

Gravitationally lensed quasars can be used as tools for a variety of astrophysical and cosmological applications \citep[e.g.][]{Treu2010,Jackson2013}, including: mapping the dark matter substructure of the lensing galaxy \citep[e.g.][]{MetcalfMadau2001,Kochanek+2004,Nierenberg+2014,Vegetti+2014,Birrer+2017}; determining the mass \citep[e.g.][]{Morgan+2010} and spin \citep{Reynolds+2014} of black holes; and measuring the properties of distant host galaxies \citep[e.g.][]{Kochanek+2001,Claeskens+2006,Peng+2006,Ding+2017}. In addition, they can also be used to constrain cosmological parameters with comparable precision to baryonic acoustic oscillation methods \citep[e.g][]{Suyu+2010,Suyu+2013} and to probe the physical properties of quasar accretion disks through microlensing studies \citep[e.g.][]{Kochanek2004,Poindexter+2008,Motta+2012}.

\cite{Ostrovski+2016} found that, for the Dark Energy Survey (DES - \citealt{DES2005,Abbott+2016}), no simulated gravitationally lensed quasar system with image separation less than 1\farcs5 is segmented into multiple catalogue sources due to limitations on survey resolution. As a result, only $23\%$ of simulated systems with pairs of quasar images are segmented into two sources. This means that to identify lensed quasars as groups of close sources of similar colours, one has to either employ 2D modelling techniques (Ostrovski et al. 2017, in prep.) or rely on cross-matching to higher resolution imaging surveys such as Gaia (FWHM $\sim$0\farcs1; \citealt{Lemon+2017}). 

Gaia \citep{Gaiamission2016} is a space observatory performing a full sky survey to study the Milky Way. Gaia data release 1 (DR1 - \citealt{GAIA2016}) contains positions for a total of $1.1\times10^{9}$ sources across the whole sky with a point source limiting magnitude of $G = 20.7$ \citep{Brown+2016}.

Here, we present the discovery of the lensed quasar PS~J0630$-$1201 from a preliminary search for gravitationally lensed quasars from Pan-STARRS1 (PS - \citealt{PS1Surveys}) combined with the Wide-field Infrared Survey Explorer (WISE - \citealt{WISE}) by cross-matching to Gaia DR1 detections. All magnitudes are quoted on the AB system. Conversions from Vega to AB for the WISE data are $W1_{AB} = W1_{Vega}+2.699$ and $W2_{AB} = W2_{Vega}+3.339$ \citep{Jarrett+2011}.

\section{Lens Discovery and Confirmation}

To select gravitationally lensed quasar candidates, we first create a \textit{morphology-independent} (i.e., not restricted to objects consistent with the point spread function) photometric quasar catalogue using Gaussian Mixture Models, a supervised machine learning method of generative classification, as described in \cite{Ostrovski+2016}. We use WISE and PS  photometry to define $g-i$, $i-W2$, $z-W1$, and $W1-W2$ colours and we employ three model classes (point sources, extended sources, and quasars) to generate a catalogue of 378,061 quasar candidates from a parent sample of 80,028,181 objects with $i$<20. To remove stellar contaminants at low galactic latitude, we apply conservative colour cuts based on the comparisons between the known quasar distribution and the distribution of our candidates as well as removal of objects with high point source probabilities. We also discard objects with a neighbouring candidate within 5$^{\prime\prime}$ yielding a total of 296,967 objects, since the spatial resolution of PS is enough to flag these systems as potential lensed quasar candidates or binary pairs.

\begin{figure}
\begin{center}
	\includegraphics[width=0.6\columnwidth]{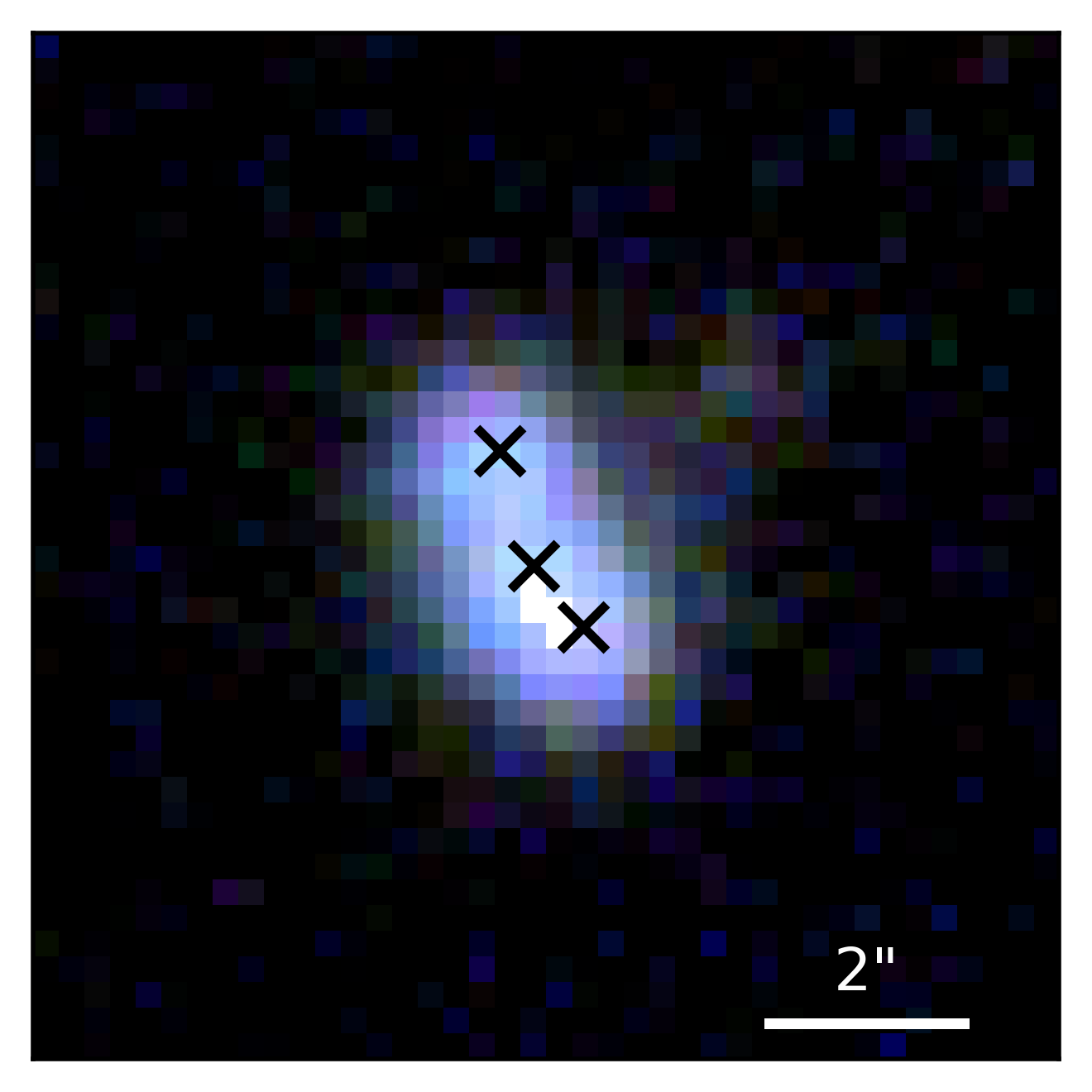}
    \caption{PS J0630$-$1201 as a $g$, $r$, $i$ PS colour composite. The single band stacked images were obtained from the PS cutout server (http://ps1images.stsci.edu/cgi-bin/ps1cutouts). The black crosses mark the positions of overlapping objects from the Gaia catalogue. North is up and East is to the left.}
    \label{fig:cutout}
\end{center}
\end{figure}

We then exploit the excellent spatial resolution and all-sky coverage of the Gaia DR1 catalogue to identify lensed quasar candidates by resolving the photometric quasars into multiple components. Gaia DR1 is known to be enormously incomplete for close-separation objects but nevertheless can identify multiple components of lensed quasars (see \citealt{Lemon+2017} for details). We cross-match our photometric quasar candidates with the Gaia DR1 catalogue using a 3$^{\prime\prime}$ search radius and find that 1,401 of the quasar candidates have 2 Gaia objects associated with their PS position, whilst 26 of the candidates have 3 Gaia associations. Visual inspection of these 26 objects revealed one lens candidate, PS J0630$-$1201, shown in Fig.~\ref{fig:cutout} with Gaia positions overlaid on the PS cutout, to be of interest. Its PS catalogue position, and PS and WISE photometry are listed in Table~\ref{tab:photometry}. The other objects with triple Gaia matches were ruled out as common contaminants, mainly single quasars with other objects nearby (that, despite being resolved in PS images, were not quasar candidates in our catalogue), as well as apparently faint interacting starbursting galaxies and likely duplicate entries in the Gaia catalogue (objects with separations of $<0\farcs1$).

\begin{table}
	\begin{center}
	\caption{PS J0630$-$1201 PanSTARRS and WISE selection photometry.}
	\label{tab:photometry}
	\begin{tabular}{|cc|}
		\hline
		RA & 06h 30m 09.11s \\
		DEC & $-$12d 01m 20.0s \\
		$g$ & 19.291 $\pm$ 0.002 \\
		$r$ & 18.566 $\pm$ 0.002 \\
		$i$ & 18.321 $\pm$ 0.001 \\
		$z$ & 17.949 $\pm$ 0.001 \\
		$Y$ & 17.873 $\pm$ 0.001 \\
		$W1$ & 16.886 $\pm$ 0.027 \\
		$W2$ & 16.643 $\pm$ 0.030 \\
		\hline
	\end{tabular}
     \end{center}
\end{table}

\subsection{Spectroscopic Follow-up at the WHT}

\begin{figure}
\begin{center}
	\includegraphics[width=1.0\columnwidth]{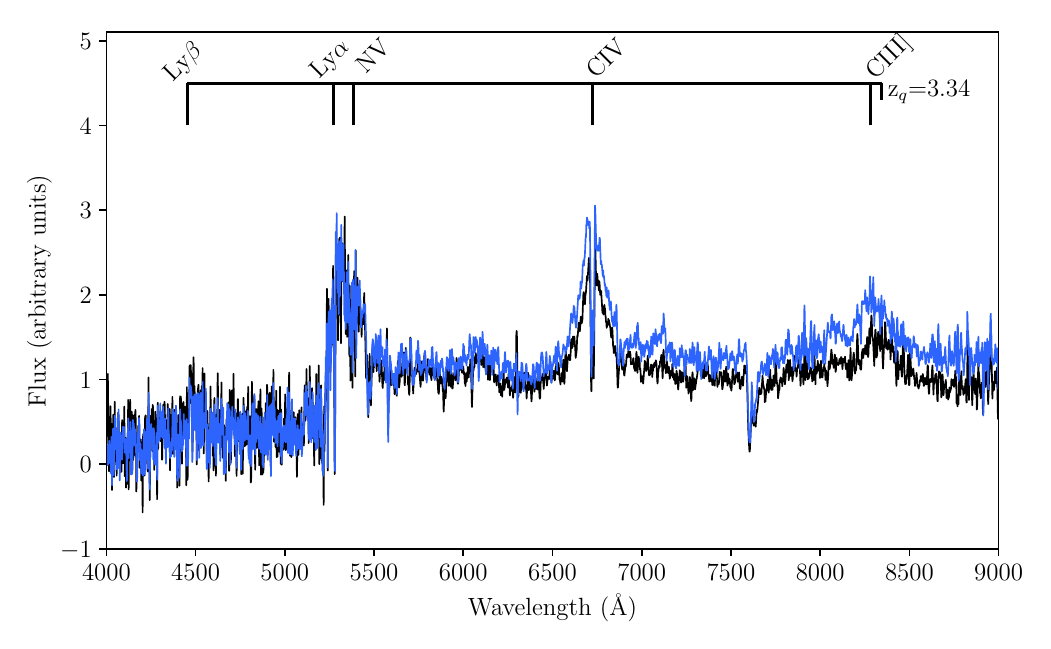}
    \caption{WHT/ISIS spectra of component C (black) and the blended components A and B (blue). On top, we have marked the location of the source quasar emission lines at $z_{q} = 3.34$.}
    \label{fig:spectra}
\end{center}
\end{figure}

Spectroscopic follow-up observations to confirm that the multiple components are multiply-imaged quasars were performed with the dual-arm Intermediate dispersion Spectrograph and Imaging System (ISIS) on the William Herschel Telescope (WHT) on the night of April 01 2017 UT. The R300B grating was employed on the blue arm, providing wavelength coverage from $\sim3200$\AA\ to $\sim$5400\AA\ with $\sim$4\AA\ resolution, and the R158R grating on the red arm resulted in coverage from $\sim$5300\AA\ to $\sim$10200\AA\ with a resolution of $\sim$7.7\AA; the 5300\AA\ dichroic was used to split the beam to the blue and red arms. Two exposures of 600s each were obtained with a slit PA of 22.5 degrees, i.e., along the direction of the three brightest images, and two clearly separated traces were visible on the red arm, with the two southern-most images not deblended, whilst no clear separation was apparent in the blue data. Extractions of these traces are shown in Figure \ref{fig:spectra}, revealing very similar spectra indicating a lensed quasar at $z_{q} = 3.34$. No features of a lower-redshift lensing galaxy are visible in the spectra.

\subsection{Adaptive Optics Follow-up with Keck}

\begin{figure*}
\begin{center}
	\includegraphics[width=2\columnwidth]{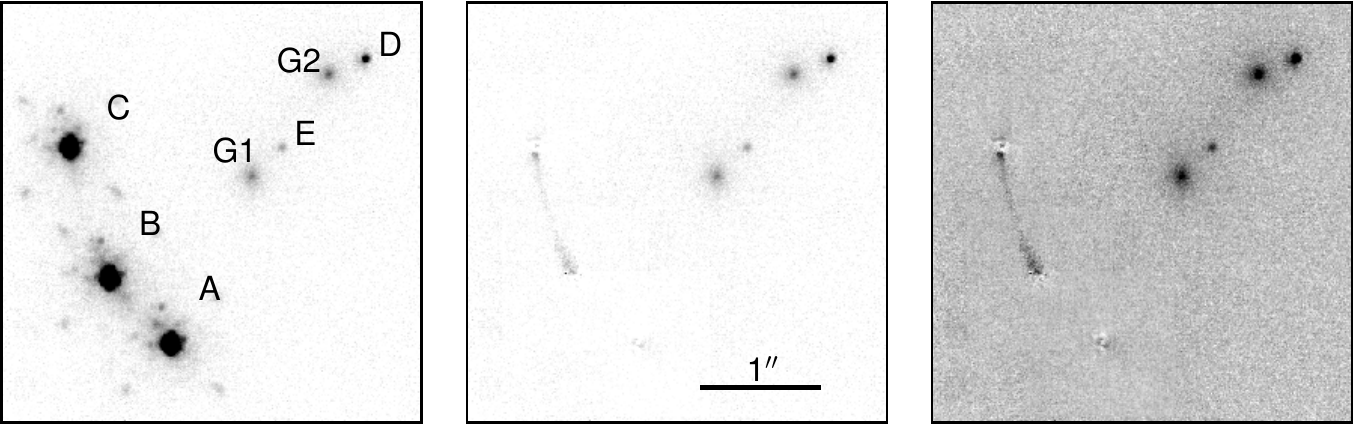}
    \caption{(Left) NIRC2 $K^{\prime}$ AO imaging of PS J0630$-$1201. The four images ABCD are in a canonical `cusp' configuration, but note the presence of two galaxies, G1 and G2, as well as the additional point source E. The additional structure around the images A, B, and C is due to poor wavefront correction. (Middle) The same as the left panel, but with a model for the AO PSF subtracted from the three brightest images, revealing the presence of a faint arc. (Right) Same as the middle but with enhanced contrast to better show the lensed quasar host galaxy.}
    \label{fig:keck}
\end{center}
\end{figure*}

After the spectroscopic confirmation of multiple quasars, the system was observed on April 11 2017 UT with the NIRC2 camera mounted on Keck II, using Keck's adaptive optics (AO) system. The NIRC2 narrow camera was used, giving a $10^{\prime\prime}\times10^{\prime\prime}$ field of view and 10mas pixels, and four 180s exposures were obtained with the $K^{\prime}$ filter. These data clearly resolve the three quasars observed with WHT spectroscopy (A, B, and C in Figure \ref{fig:keck}), and also reveal two additional point-like objects (D and E) and two extended objects (G1 and G2). Note that most of the structure around the bright images is an artefact (``waffling'') due to AO correction problems with the low-bandwidth wavefront sensor.

The PSF of image C appears to have a structure extending down from the core of the PSF that is not seen in images A or B but could be consistent with a lensed arc. We therefore produced a pixellated model of the PSF around the images ABC to remove it and increase the dynamic range of the image. In the first iteration of this procedure an arc between images B and C was clearly visible, and we therefore re-fitted for the PSF excluding pixels that contain arc flux. The residuals of this fit are shown in the middle and right panels of Figure \ref{fig:keck}.

\section{Lens Mass Modelling}
\label{massmod}

The AO imaging data show that PS J0630$-$1201 is a lensed quasar in a `cusp' configuration, with two lensing galaxies. The presence of a fifth point source, E in Figure \ref{fig:keck}, is intriguing as it is located approximately where the fifth (demagnified) image would be expected to appear. To understand the nature of E as a possible fifth quasar image, we initially create a mass model using the positions of the brighter quasar images (A to D) and the two lensing galaxies, and use this model to predict the location of any additional images. We first determine the position of each point source by modelling them with Gaussian or Moffat profiles, and we fit Sersic profiles to G1 and G2. We also perform moment-based centroiding in a range of apertures, and use the spread of all of our measurements to estimate the uncertainties on the positions; these are typically $\sim$1~mas for the point sources and $\sim$10 to 20~mas for the galaxies. However, we also find that the relative positions of the quasar images change from exposure to exposure, presumably due to atmospheric fluctuations, and we therefore impose a 5~mas minimum uncertainty on each position. We simultaneously fit the PS imaging data ($grizY$) whilst fitting the Keck data ($K^{\prime}$), and our photometric and astrometric results are given in Table \ref{tab:compphotometry}.

\begin{table*}
        \centering
        \caption{Relative astrometry and photometry of lens components. $grizYK^{\prime}$ of D and E are combined with G2 and G1 respectively due to blending.}
    \vspace*{2mm}
        \label{tab:compphotometry}
        \begin{threeparttable}
        \resizebox{\textwidth}{!}{
        \begin{tabular}{cccccccccc} 
                \hline 
                 & $G^{\dag}$ & $g$ & $r$ & $i$ & $z$ & $Y$ & $K^{\prime}$ & $\Delta\alpha$ & $\Delta\rm{\delta}$\\
                \hline
                A & 19.95$\pm$0.01 & 20.65$\pm$0.05 & 19.92$\pm$0.04 & 19.65$\pm$0.04 & 19.27$\pm$0.01 & 19.22$\pm$0.05 & 18.65$\pm$0.01 & 0.674 $\pm$ 0.005 & $-$1.421 $\pm$ 0.005 \\
                B & 19.76$\pm$0.01 & 20.54$\pm$0.05 & 19.73$\pm$0.05 & 19.80$\pm$0.03 & 19.18$\pm$0.02 & 19.23$\pm$0.04 & 18.49$\pm$0.01 & 1.192 $\pm$ 0.005 & $-$0.865 $\pm$ 0.005  \\
                C & 19.61$\pm$0.01 & 20.53$\pm$0.05 & 19.70$\pm$0.05 & 19.66$\pm$0.04 & 19.24$\pm$0.02 & 19.15$\pm$0.04 & 18.66$\pm$0.01 & 1.530 $\pm$ 0.005 & 0.248 $\pm$ 0.005 \\
        D(+G2) & $-$ & 23.03$\pm$0.16 & 21.80$\pm$0.05 & 21.59$\pm$0.04 & 21.32$\pm$0.04 & 20.85$\pm$0.03 & 21.30$\pm$0.01 &  $-$0.966 $\pm$ 0.005 & 0.994 $\pm$ 0.005 \\
        E(+G1) & $-$ & 23.33$\pm$0.34 & 23.16$\pm$0.14 & 21.93$\pm$0.09 & 21.83$\pm$0.09 & 21.60$\pm$0.09 & 22.76$\pm$0.01 &  $-$0.257 $\pm$ 0.005 & 0.241 $\pm$ 0.005 \\
        G1 & $-$ & $-$ & $-$ & $-$ & $-$ & $-$ & 20.86$\pm$0.01 & 0.000 $\pm$ 0.013 & 0.000 $\pm$ 0.010 \\ 
        G2 & $-$ & $-$ & $-$ & $-$ & $-$ & $-$ & 20.93$\pm$0.01 & $-$0.650 $\pm$ 0.017 & 0.862 $\pm$ 0.005 \\
                \hline
        \end{tabular}}
	\begin{tablenotes}
			\item $^{\dag}$ Gaia DR1 catalogue $G$-band photometry for components A,B, and C.
    \end{tablenotes}
    \end{threeparttable} 
\end{table*}

We use the positions of the galaxies and point sources to constrain a lensing mass model. The two galaxies are initially modelled as singular isothermal spheres and we use the positions of the four brighter images to infer their Einstein radii using the \texttt{lensmodel} software \citep{GRAVLENS}. The best-fit lens model predicts a fifth image near the location of image E with a flux approximately half of image D, comparable to the observed flux ratio. We consequently use the position of E to constrain a more realistic lens model, allowing the two galaxies to have some ellipticity, and we include an external shear. This model has five free parameters for each galaxy (two position parameters, an Einstein radius, and an ellipticity with its orientation) and two additional parameters each for the position of the source and the external shear, i.e., there are 14 total parameters. We likewise have 14 constraints from the observed positions of the five quasar images and the two lensing galaxies.

Using image plane modelling, we sample the parameter space using \texttt{emcee} \citep{foremanmackey2013} and, as expected, find a best-fit model with ${\chi}^{2} \sim 0$. Both mass components are inferred to be coincident with the light, with Einstein radii for G1 and G2 of 1\farcs01$\pm$0.01 and 0\farcs58$\pm$0.01, respectively. Both masses are also mildly flattened, but we find that the light and mass \textit{orientations}, given in Table \ref{tab:J0630_shapes}, are misaligned, significantly so for G2. We also note that the shear is well constrained ($\gamma=0.14\pm0.01$ with PA $-76$) and is particularly large, though there is no nearby galaxy along the shear direction. The total magnification for the system is $\sim$53.

\begin{table}
	\begin{center}
	\caption{Light and mass shape parameters (position angle, PA, and axis ratio, q) for the lens galaxies.}
	\label{tab:J0630_shapes}
	\begin{threeparttable}
	\begin{tabular}{ccccc}
  \hline
   & PA (light) & $q$ (light) & PA (mass) & $q$ (mass)\\
  \hline
       G1 &   $-$9 $\pm$ 4 & 0.81 $\pm$ 0.03 & $-$21 $\pm$ 6 & 0.81 $\pm$ 0.02 \\
       G2 &   $-$69 $\pm$ 4 & 0.79 $\pm$ 0.03 & $-$2 $\pm$ 7 & 0.83 $\pm$ 0.04 \\
  \hline
	\end{tabular}
    \end{threeparttable}
     \end{center}
\end{table}

\section{Discussion and conclusions}

We have presented spectroscopic and imaging data that confirm that PS~J0630$-$1201 is a quasar at $z_{q} = 3.34$ lensed into five images by two lensing galaxies. We are able to fit the positions of the five images well with a two-SIE lens model and  recover flux ratios to within 30\% with discrepancies likely caused by microlensing and/or differential extinction and reddening, as evidenced by the strongly varying flux ratio between images B and C from optical to near-infrared wavelengths. However, we find that for both lenses the ellipticity of the mass is not consistent with the ellipticity of the galaxy light (Table \ref{tab:J0630_shapes}) and the inferred shear is quite large. This could be the result of an additional mass component, e.g., a dark matter halo that is not coincident with either galaxy \citep[e.g.,][]{Shu+2016}. In that case, the weak demagnification of the fifth image might indicate that the dark matter halo is not cuspy \citep[e.g.,][]{Collett+2017}, although constraints from the quasar image positions alone are not sufficient to test this. Deeper imaging of the arc of the lensed host galaxy and observations at radio wavelengths, where extinction and the effects of microlensing are no longer important, will help to constrain a more complex model for the mass distribution.

The relatively bright fifth image of PS~J0630$-$1201 also presents the possibility of obtaining four new time delay measurements, for a total of ten time delays. Based upon our current best lens model, these delays should range between 1 and 245 days assuming that the lens redshifts are $z\sim1$ (Table~\ref{tab:tdelaymatrix}). Because of the overall compactness of the system and the presence of the two lensing galaxies, it would be difficult to obtain time delays from the fifth image with conventional seeing-limited monitoring programmes. However, if such a campaign observed a sudden brightening (or dimming) event in one of the brighter images, dedicated monitoring with a high-resolution facility \citep[e.g., Robo-AO;][]{Baranec+2014} could yield an observation of the delayed brightening of the fifth image.

Several other lensed quasars have been observed to have bright additional images, and these systems typically also have multiple lensing galaxies \citep[with the exception being the radio-loud double PMN~J1632-0033, which has an observed highly de-magnified central image][]{Winn+2004}. The cluster lens SDSS~J1004+4112 \citep{Inada+2005} is a quad with an observed `central' image, but the the complexity of the mass distribution makes it very difficult to estimate time delays \citep{Fohlmeister+2007}. PMN~J0134-0931 is a radio-loud quasar that is being lensed into five images by two galaxies \citep{Winn+2003}. However, the image separations are very small so measuring time delays -- expected to range from an hour to two weeks \citep{Keeton+2003} -- will be extremely difficult. B1359+154 \citep{Rusin+2001} and SDSS~J2222+2745 \citep{Dahle+2013} are six-image lens systems, and in both cases the lens is a compact group of three galaxies. These lenses have intrinsically more complex mass distributions, but time delays may still be informative for cosmography. However, most of the independent time delays are expected to be less than a day for B1359+154 \citep{Rusin+2001} and quite long (400 to 700 days) for SDSS~J2222+2745 \citep{Sharon+2017}. PS~J0630$-$1201 therefore appears to be the most promising lens for measuring additional independent time delays.

\begin{table}
	\centering
	\caption{Predicted time delays between the image pairs in PS J0630$-$1201. All values are in days. The values in the top right (bottom left) are for a lens redshift of 1 (0.5). Light arrives in the images in the following order: CABED}
    \vspace*{5mm}
	\label{tab:tdelaymatrix}
	\begin{tabular}{cccccc} 
		\hline
		 & A & B & C & D & E \\
		\hline
		A & $-$ & 0.9 & 1.8 & 243 & 208 \\
		B & 0.4 & $-$ & 2.7 & 242 & 207 \\
		C & 0.7 & 1.1 & $-$ & 245 & 210 \\
		D & 97.8 & 97.4 & 98.5 & $-$ & 35 \\
		E & 83.7 & 83.3 & 84.4 & 14.1 & $-$ \\
		\hline
	\end{tabular}
\end{table}

\section*{Acknowledgements}

We are grateful to the STRIDES collaboration for many useful discussions about quasar lens finding. FO, CAL, RGM, and SLR acknowledge the support of the UK Science and Technology research Council (STFC), and MWA also acknowledges STFC support in the form of an Ernest Rutherford Fellowship. FO acknowledges the current support of the Conselho Nacional de Desenvolvimento Científico e Tecnológico (CNPq - grant number 150151/2017-9) and was supported jointly by CAPES (the Science without Borders programme) and the Cambridge Commonwealth Trust during part of this research. CDF and CER acknowledge support from the US National Science Foundation through grant number AST-1312329. 

The Pan-STARRS1 Surveys (PS1) and the PS1 public science archive have been made possible through contributions by the Institute for Astronomy, the University of Hawaii, the Pan-STARRS Project Office, the Max-Planck Society and its participating institutes, the Max Planck Institute for Astronomy, Heidelberg and the Max Planck Institute for Extraterrestrial Physics, Garching, The Johns Hopkins University, Durham University, the University of Edinburgh, the Queen's University Belfast, the Harvard-Smithsonian Center for Astrophysics, the Las Cumbres Observatory Global Telescope Network Incorporated, the National Central University of Taiwan, the Space Telescope Science Institute, the National Aeronautics and Space Administration under Grant No. NNX08AR22G issued through the Planetary Science Division of the NASA Science Mission Directorate, the National Science Foundation Grant No. AST-1238877, the University of Maryland, Eotvos Lorand University (ELTE), the Los Alamos National Laboratory, and the Gordon and Betty Moore Foundation.

This publication makes use of data products from the Wide-field Infrared Survey Explorer, which is a joint project of the University of California, Los Angeles, and the Jet Propulsion Laboratory/California Institute of Technology, funded by the National Aeronautics and Space Administration.

This work has made use of data from the European Space Agency (ESA) mission Gaia (\url{https://www.cosmos.esa.int/gaia}), processed by the Gaia Data Processing and Analysis Consortium (DPAC, \url{https://www.cosmos.esa.int/web/gaia/dpac/consortium}). Funding for the DPAC has been provided by national institutions, in particular the institutions participating in the Gaia Multilateral Agreement.

The William Herschel Telescope is operated on the island of La Palma by the Isaac Newton Group of Telescopes in the Spanish Observatorio del Roque de los Muchachos of the Instituto de Astrofísica de Canarias.

Some of the data presented herein were obtained at the W.M. Keck Observatory, which is operated as a scientific partnership among the California Institute of Technology, the University of California and the National Aeronautics and Space Administration. The Observatory was made possible by the generous financial support of the W.M. Keck Foundation. The authors wish to recognize and acknowledge the very significant cultural role and reverence that the summit of Mauna Kea has always had within the indigenous Hawaiian community.  We are most fortunate to have the opportunity to conduct observations from this mountain.

This research made use of Astropy, a community-developed core Python package for Astronomy (Astropy Collaboration, 2013).


\bibliographystyle{mnras/mnras}
\bibliography{/Users/fernanda/Dropbox/Papers/PS1J0630/rev/rev2/references}


\bsp	
\label{lastpage}
\end{document}